\documentclass[preprint,number,12pt]{elsarticle} 
\usepackage{graphicx}
\usepackage{color} 
\usepackage{epsfig}
\usepackage{times}
 
\begin{document}
\begin{frontmatter}
\title{Alternating-Order Interpolation in a Charge-Conserving Scheme for 
Particle-In-Cell Simulations}
\author{Igor V. Sokolov}
\address{Center for Radiative Shock Hydrodynamics, University of
  Michigan, 2455 Hayward Str., Ann Arbor MI48109; igorsok@umich.edu}
\begin{abstract}
We discuss the interpolation of the electric and magnetic fields within
a charge-conserving Particle-In-Cell scheme. The choice of the interpolation procedure for the fields
acting on a particle can be constrained by analyzing
conservation of the energy and the particle
generalized momentum. The better conservative properties are achieved,
if the alternating-order form-factor is used for interpolation, which
combines the lower-order and higher-order interpolation from integer
and semi-integer points of a staggered grid. This approach allows us to 
significantly 
improve both the results quality and the computational efficiency for the charge conserving scheme.  
\end{abstract}
\end{frontmatter}

\section{Introduction: charge-conserving PIC schemes.}
Here we discuss the conservative properties of the Particle-In-Cell
(PIC) numerical schemes. Recall that usually the conservative schemes
are  employed to solve the system of conservation laws. For example, the
continuity equation, 
$\frac{\partial \rho_m}{\partial t}+
\nabla\cdot(\rho_{m}{\bf u})=0$, for the fluid mass density, $\rho_m$, may be advanced 
through the time step, $\Delta t$, using the conservative scheme:
$V_i(\rho_m)_i^{n+1}=V_i(\rho_m)_i^{n}-\Delta t\sum_j{\sigma_{ij}}$,
with the computational domain split into a set of {\it
  control volumes} (``cells''), the mass density at a given
time instant, $t^n=n\Delta t$, averaged over
the volume of the cell, $i$, and the mass flux through the $ij$ face
averaged over the time step: 
\begin{equation}\label{gencons}
(\rho_m)_i^{n}=\frac1{V_i}\int{(\rho_m)_{t=t^n}dV_i},\qquad\sigma_{ij}=\frac1{\Delta t}\int_{t^n}^{t^{n+1}}{dt\int{\left(d{\bf S}_{ij}\cdot{\bf u}\rho_m\right)}},
\end{equation} 
where ${\bf u}$ is the fluid velocity and the face area vector, 
${\bf S}_{ij}$ is directed from 
cell $i$, to cell $j$. Since
$\sigma_{ij}=-\sigma_{ji}$, 
the total mass is conserved: $\sum_i{(\rho_m)_i^{n+1}}=\sum_i{(\rho_m)_i^{n}}$. 
  
This idea may be employed to assure the charge
conservation within the PIC method, although
the governing equations differ from conservation laws:
\begin{equation}\label{1}
 \frac{d{\bf w}_p}{dt}=\frac{q_p}{m_p}\left({\bf E}({\bf x}_p)+
\left[\frac{{\bf u}_p}c\times {\bf B}({\bf x}_p)\right]\right),\qquad 
\frac{{\bf u}_p}c={\bf w}_p/\sqrt{({\bf w}_p)^2+c^2},
\end{equation}
\begin{equation}\label{second_pair}
\frac{\partial {\bf E}}{\partial t} =-4\pi{\bf J}+c[\nabla\times{\bf B}],\qquad 
(\nabla\cdot{\bf E})=4\pi\rho,
\end{equation}
\begin{equation}
\frac{\partial {\bf B}}{\partial t} =-c[\nabla\times{\bf E}],\qquad
(\nabla\cdot{\bf B})=0,
\end{equation}
\begin{equation}\label{coorddepend}
\frac {d{\bf x}_p}{dt}={\bf u}_p,\qquad
 \rho=\sum_p{q_p\delta({\bf x}-{\bf x}_{p})},\qquad {\bf J}=\sum_p{q_p{\bf u}_p\delta({\bf x}-{\bf x}_{p})},
\end{equation}
\begin{equation}\label{vectorpot}
 {\bf E}=-\frac1c\frac{\partial {\bf A}}{\partial t},\qquad  {\bf B}=[\nabla\times{\bf A}],
\end{equation}
where the index $p$ enumerates particles (electrons, ions), $ {\bf
  w}_p$ is a 
momentum-to-mass ratio, and other notations are usual. 
The particle charge density, $\rho$, obeys not only the
continuity equation, $\frac{\partial \rho}{\partial t}+\nabla\cdot{\bf
  J}=0$, but also the Poisson equation. Therefore, the charge
  conservation property is formulated as the relationship between the  
charge density and the electric 
field,  ${\bf E}$, and/or the electric current, ${\bf J}$. 

In the present paper we discuss only the CHarge-Conserving PIC (ChCPIC)
schemes. The need to introduce this family of numerical schemes is 
that to simulate fully relativistic motions of plasmas in 
strong rapidly varying electromagnetic fields (such as the laser field, 
for example), one may want to directly advance the 
electric field using the first Eqs.(\ref{second_pair}). The 
``charge conservation'' property in this case allows one to fulfill the Poisson 
equation for this advanced field without solving the equation directly. The way to assure that the charge conservation was developed
in \cite{morse,buneman1,buneman2,eastwood,eastwood2} (see also
\cite{esirkepov,othmer,umeda,barth} and references therein). First, a 
staggered grid should be used as in \cite{yee} to ensure the finite-difference approximations for 
$[\nabla\times{\bf B}]$, $[\nabla\times{\bf E}]$ to be
divergence-free, as we discuss briefly in Sec. 2. Second, the
currents through the cell faces should be calculated in such way that their
divergence balances  the charge leakage from the cell. In Sec. 3, 
this is done using the virtual path integration to solve the time integral in
Eq.(\ref{gencons}). However, once the way to compute the particle
current is modified, we should also modify accordingly the
scheme for interpolating electric and magnetic fields acting on this
particle, which is the goal of this paper.

Note an important distinction  of the ChCPIC schemes from the general
cloud-in-cell scheme, which is explicitly pronounced once we follow the
conservative schemes ideology. On one hand, a ``cloud''  within
the framework of the cloud-in-cell scheme can be thought of as some
charge density distribution, $\rho_p({\bf x},{\bf x}_p)$, centered about
the particle coordinate vector, ${\bf x}_p$. Following this ideology, the {\it
  point-wise values} of the shape-function (form-factor), $\rho_p({\bf
  x},{\bf x}_p)$, at the points ${\bf x}$, where the electromagnetic
fields are localized, should be used then as the {\it interpolation
  coefficients} to average the electromagnetic fields acting on the
particle, and to calculate the Lorentz force. Contrarily, the
conservative schemes approach assumes that the cloud charge density
should be {\it integrated} over some control volume or its faces, to
represent the contribution to the plasma density and the plasma
electric current (see \ref{gencons}), thus making inapplicable the form-factor
point-wise values. To constrain the field interpolation procedure, in Sec. 
4 we discuss the accuracy of conservation, for the
energy integral and for the particle generalized momentum. Their
conservation in the ChCPIC scheme can be
achieved, if the alternating-order form-factor is employed in the
interpolation procedure. This alternating-order form-factor
combines the interpolation {\it of different order} for different
physical variables to interpolate. Equivalently, it integrates
the ``cloud'' charge density over {\it faces} and {\it edges} of the
control volume to interpolate face-centered and edge-centered 
electromagnetic field values.  
\section{Grid geometry and notations.}\label{grid}
We use a 3D Cartesian grid in the domain  $0\le x \le L_x$, $0\le y \le L_y$, $0\le z \le  L_z$, 
split for  $N_x* N_y* N_z$ cells.
The coordinates of the cell corners are $(i\ \Delta x,j\ \Delta y,k\ \Delta z)$, where 
$i$, $j$, $k$ are integers and $ \Delta x=L_x/N_x$, $\Delta y=L_y/N_y$, 
$\Delta z=L_z/N_z$  are the cell sizes.  
Then we introduce the normalized coordinates,  
$\tilde{x}=x/\Delta x,\ \tilde{y}=y/\Delta y,\ \tilde{z}=z/\Delta z$,
and time, $\tilde{t}=t/\Delta t$, and use
them below with no tilde. Magnetic field, electric current, 
and particle momenta are defined at semi-integer time instants, $t=n+1/2$, the
electric field and the particle coordinates - at integer time instants, $t=n$. 
In the normalized coordinates, Eqs.(\ref{coorddepend})
read: 
\begin{equation}\label{coorddimless}
\frac{d{\bf x}_p}{dt}=\frac{{\bf u}_p}{c}\cdot{\bf
  diag}(c_x,c_y,c_z),\quad 
\rho=\sum_p{\frac{q_p}V\delta({\bf x}-{\bf x}_p)},
\end{equation} 
where $V=\Delta x \Delta y \Delta z$ and $c_x=c\Delta t/\Delta x,...$. The 
{\it grid functions} are defined: the cell-centered charge density, 
$\rho_{i+1/2,j+1/2,k+1/2}$; the
electric field, $E^{(x)}_{i,j+1/2,k+1/2}$, $E^{(y)}_{i+1/2,j,k+1/2}$, 
$E^{(z)}_{i+1/2,j+1/2,k}$, and the current density components, $J^{(x)}_{i,j+1/2,k+1/2}$, 
$J^{(y)}_{i+1/2,j,k+1/2}$, $J^{(z)}_{i+1/2,j+1/2,k}$, 
defined at the centers of the faces, normal to the axis, $x$, $y$, $z$;
and the
magnetic field 
components, $B^{(x)}_{i+1/2,j,k}$, $B^{(y)}_{i,j+1/2,k}$, 
$B^{(z)}_{i,j,k+1/2}$, defined at the midpoints of the edges directed along the 
axis, $x$, $y$, $z$. The subscript indexes denote  
coordinates of the point at which the grid
function is defined. 

The PIC scheme as taken from \cite{birdsall85} with the suggested
modifications is described in the
Appendix. The 
algorithm involves interpolation for the fields acting on
a particle:
$$
E^{(x)}({\bf x}^{n}_p)=\sum_{i,j,k}{\alpha^{(x)}_{i,j+1/2,k+1/2}({\bf x}^{n}_p)E^{(x)}_{i,j+1/2,k+1/2}},
$$
$$
E^{(y)}({\bf x}^{n}_p)=\sum_{i,j,k}{\alpha^{(y)}_{i+1/2,j,k+1/2}({\bf x}^{n}_p)E^{(y)}_{i+1/2,j,k+1/2}},
$$
$$
E^{(z)}({\bf x}^{n}_p)=\sum_{i,j,k}{\alpha^{(z)}_{i+1/2,j+1/2,k}({\bf x}^{n}_p)E^{(z)}_{i+1/2,j+1/2,k}},
$$
$$
B^{(x)}({\bf x}^{n}_p)=\sum_{i,j,k}{\beta^{(x)}_{i+1/2,j,k}({\bf x}^{n}_p)B^{(x)}_{i+1/2,j,k}},
$$
$$
B^{(y)}({\bf x}^{n}_p)=\sum_{i,j,k}{\beta^{(y)}_{i,j+1/2,k}({\bf x}^{n}_p)B^{(y)}_{i,j+1/2,k}},
$$
$$
B^{(z)}({\bf x}^{n}_p)=\sum_{i,j,k}{\beta^{(z)}_{i,j,k+1/2}({\bf x}^{n}_p)B^{(z)}_{i,j,k+1/2}},
$$
where $\alpha$, $\beta$ are the weights, their sums for a
given particle should be equal to one:
\begin{equation}\label{one}
\sum_{i,j,k}{\alpha^{(x)}_{i,j+1/2,k+1/2}({\bf x}^{n}_p)}=1,...\qquad 
\sum_{i,j,k}{\beta^{(x)}_{i+1/2,j,k}({\bf x}^{n}_p)}=1,...
\end{equation}
Herewith, we provide only the 
expressions for $x$ component of vectors, 
whenever possible, denoting the generalization for the other components by '...'. The
contribution to the current density from a charged
particle can be expressed in terms of the particle position, ${\bf
  x}^{n}_p$, and its velocity, ${\bf u}^{n+1/2}_p$, and/or 
${\bf x}^{n+1}_p$: 
$$
J^{(x)\ n+1/2}_{i,j+1/2,k+1/2}=\sum_{p}{\frac{q_p}V\xi^{(x)}_{i,j+1/2,k+1/2}({\bf x}^{n}_p,\ {\bf x}^{n+1}_p)},
$$
$$
J^{(y)\ n+1/2}_{i+1/2,j,k+1/2}=\sum_{p}{\frac{q_p}V\xi^{(y)}_{i+1/2,j,k+1/2}({\bf x}^{n}_p,\ {\bf x}^{n+1}_p)},
$$
\begin{equation}\label{xi}
  J^{(z)\ n+1/2}_{i+1/2,j+1/2,k}=\sum_{p}{\frac{q_p}V\xi^{(z)}_{i+1/2,j+1/2,k}({\bf x}^{n}_p,\ {\bf x}^{n+1}_p)}.
\end{equation}
The advantage of the staggered grid is that the  magnetic field
divergence  
does not change and equals zero as long as it is initially equal to
zero. Analogously, $[\nabla\times{\bf B}]$ term does
not affect the electric field divergence. The
Poisson equation,
$4\pi\rho^{n}_{i+1/2,j+1/2,k+1/2}=(E^{n\ (x)}_{i+1,j+1/2,k+1/2}-E^{n\ (x)}_{i,j+1/2,k+1/2})/\Delta
x+...$, is satisfied, if: 
\begin{eqnarray}\label{chargelaw}
\frac{\rho^{n}_{i+1/2,j+1/2,k+1/2}-\rho^{n+1}_{i+1/2,j+1/2,k+1/2}}{\Delta t}=
\frac{J^{(x)\ n+1/2}_{i+1,j+1/2,k+1/2}-J^{(x)\ n+1/2}_{i,j+1/2,k+1/2}}{\Delta
  x}+\nonumber\\
+\frac{J^{(y)\ n+1/2}_{i+1/2,j+1,k+1/2}-J^{(y)\ n+1/2}_{i+1/2,j,k+1/2}}{\Delta
  y}+\frac{J^{(z)\ n+1/2}_{i+1/2,j+1/2,k+1}-J^{(z)\ n+1/2}_{i+1/2,j+1/2,k}}{\Delta
  x}.
\end{eqnarray}
\section{Charge density and charge conservation law.}
\subsection{Form-factors.}
To discretize the charge and current densities, one
needs to specify the numerical representation for $\delta$ functions in 
Eqs.(\ref{coorddimless}). We do this using the family of form-factor
functions, $f^{(l)}(x,x_p)$, where the form-factor of a zero order 
is a cap-function: $f^{(0)}(x,x_p)=1$, 
and the higher-order form-factors are recursively defined: 
$
f^{(l+1)}(x,x_p)=\int_{x-1/2}^{x+1/2}{f^{(l)}(x^\prime,x_p)dx^\prime}$.
All form-factors: (1) are symmetric functions of $x-x_p$; (2) turn to
zero at $|x-x_p|>(l+1)/2$; and (3):
$$\frac{\partial f^{(l+1)}(x,x_p)}{\partial x}= -\frac{\partial f^{(l+1)}(x,x_p)}
{\partial x_p}=f^{(l)}(x+1/2,x_p)-f^{(l)}(x-1/2,x_p).$$

We are interested both in point values of the form-factor
function, and in its integrals over the grid size. So, for a chosen form-factor,
$f(x,x_p)=f^{(l)}(x,x_p)$, we introduce: 
$$f_i(x_p)=f(i,x_p),\qquad   
F_i(x_p)=\int_{-\infty}^i{f(x^\prime,x_p)dx^\prime}$$ 
and 
\begin{equation}\label{intff}
\Delta F_{i+1/2}(x_p)=F_{i+1}(x_p)-F_i(x_p)=\int_{i}^{i+1}{f(x^\prime,x_p)dx^\prime}.
\end{equation}
By definition, 
$
\Delta F_{i+1/2}(x_p)=
f^{(l+1)}(i+1/2,x_p)$. 
 The applicability of the form-factors 
for constructing the interpolation weights, which should satisfy
Eq.(\ref{one}), is ensured by the equation: 
\begin{equation}\label{eq:sum1}
\sum_{i}{f^{(l)}(x+i,x_p)}=\int_{|x-x_p|\le l/2}{f^{(l-1)}(x,x_p)dx}=1.
\end{equation}
From Eq.(\ref{eq:sum1}) we can obtain yet another identity to be used
below:
\begin{equation}\label{eq:sumcurrent}
\sum_i{\left[F_i(x_p)-F_i(x_p^\prime)\right]}=x_p^\prime-x_p.
\end{equation}
To prove Eq.(\ref{eq:sumcurrent}) one can note that for $x_p=x_p^\prime$
both sides of the equation turn to zero and that on taking the derivative
of this equation over $x_p^\prime$ we obtain the earlier proven
Eq.(\ref{eq:sum1}).
\subsection{Conservative scheme for electric charge}
A particle can be thought of as a cloud with the charge 
density, $\rho_p(x,y,z)=q_pf(x,x_p)f(y,y_p)f(z,z_p)/V$. Following the conservative
scheme idea, we define the contribution from this particle to the
charge density {\it grid function} not as a point value of $\rho_p$ in
the cell centers, but via 
Eqs.(\ref{gencons},\ref{intff}):
\begin{equation}\label{rho}
\rho^{n}_{i+1/2,j+1/2,k+1/2}=\frac1V\sum_p{q_p\Delta F_{i+1/2}(x^{n}_p)
\Delta F_{j+1/2}(y^{n}_p)\Delta F_{k+1/2}(z^{n}_p)}.
\end{equation}
Note that the {\it integrated} over the cell size form-factor value, $\Delta F_{i+1/2}(x^{n}_p),...$ is
at the same time the {\it point-wise} value of the form-factor function
of by unity higher order, $\Delta
F_{i+1/2}(x^{n}_p)=f^{(l+1)}_{i+1/2}(x^{n}_p)$. 

Assuming that within the time interval, $(n,n+1)$, the particle
moves from the point 
${\bf x}^{n}_p$ to the 
point ${\bf x}^{n+1}_p$ along an arbitrary {\it
  virtual} path 
${\bf x}_v(t)$,  we define the particle currents, 
$\xi^{(x,y,z)}$, in Eq.(\ref{xi}) following Eqs.(\ref{gencons},\ref{intff},\ref{rho})):
$$
\xi^{(x)\ n+1/2}_{i,j+1/2,k+1/2}({\bf x}^{n}_p,{\bf x}^{n+1}_p)
=\int\limits_{n}^{n+1}\int\limits_{j}^{j+1}\int\limits_{k}^{k+1}\frac{\Delta x}{\Delta
  t}\frac{dx_v}{dt}
f_i(x_v)
f(y^\prime,y_v)
f(z^\prime,z_v)dz^\prime dy^\prime dt=
$$
\begin{equation}\label{Jx}
=-\frac{\Delta x}{\Delta t}\int\limits_{n}^{n+1}{\frac{dF_i(x_v)}{dt}
\Delta F_{j+1/2}(y_v)\Delta F_{k+1/2}(z_v)dt},
\end{equation}
 \begin{equation}
\xi^{(y)\ n+1/2}_{i+1/2,j,k+1/2}({\bf x}^{n}_p,{\bf x}^{n+1}_p)=
-\frac{\Delta y}{\Delta t}\int\limits_{n}^{n+1}{\frac{dF_j(y_v)}{dt}\Delta F_{i+1/2}(x_v)\Delta F_{k+1/2}(z_v)dt},
\end{equation}
\begin{equation}\label{Jz}
\xi^{(z)\ n+1/2}_{i+1/2,j+1/2,k}({\bf x}^{n}_p,{\bf
  x}^{n+1}_p)=-\frac{\Delta z}{\Delta t}\int\limits_{n}^{n+1}{
\frac{dF_k(z_v)}{dt}\Delta F_{i+1/2}(x_v)\Delta F_{j+1/2}(y_v)dt}.
\end{equation}
By definition, Eqs.(\ref{rho}-\ref{Jz}) 
satisfy the charge conservation law as in Eq.(\ref{chargelaw}). This
can be verified by observing that 
the linear combination of Eqs.(\ref{Jx}-\ref{Jz}) as presented in
Eq.(\ref{chargelaw}), reduces to $\int{\frac{d}{dt}[\Delta
F_{i+1/2}(x_v)\Delta F_{j+1/2}(y_v)\Delta F_{k+1/2}(z_v)]dt}$.

Discuss a choice of the virtual path. For the straight path,  
${\bf x}_v(t)={\bf x}^{n}_p+(t-n){\bf u}^{n+1/2}_p\cdot{\bf
  diag}(c_x,c_y,c_z)/c$, the integrands in
Eqs.(\ref{Jx}-\ref{Jz}) are piecewise polynomials of the order of
$3l+2$. The integration was performed in \cite{buneman1},\cite{eastwood2} 
only for the lowest order form-factor, $l=0$. With a finite time step, 
especially for higher-order form-factors, the integration becomes 
sophisticated, because the form-factor is a piecewise smooth function. All 
ranges of smoothness should be accounted for separately while 
evaluating the integral analytically.  

To overcome this difficulty, we observe that the
integrals in Eqs.(\ref{Jx}-\ref{Jz}) can be solved, if the
virtual path is composed of
the edges of the rectangular box, such that the points ${\bf x}_p^{(n)}$, 
${\bf x}_p^{(n+1)}$ are the opposite corners of this box and its edges
are parallel to the coordinate axis. Upon calculating the integral in
Eq.(\ref{Jx}) as a sixth of a sum of the integrals along six 
possible virtual paths, we find:
$$
\xi^{(x)\ n+1/2}_{i,j+1/2,k+1/2}({\bf x}^{n}_p,{\bf x}^{n+1}_p)=
-\frac16\frac{\Delta x}{\Delta t}\left[F_i(x_p^{n+1})-F_i(x_p^n)\right]\times
$$
$$
\times\{2\left[\Delta F_{j+1/2}(y_p^{n+1})\Delta
  F_{k+1/2}(z_p^{n+1})+\Delta F_{j+1/2}(y_p^{n})\Delta
  F_{k+1/2}(z_p^{n})\right]+
$$
$$
+\Delta F_{j+1/2}(y_p^{n+1})\Delta
  F_{k+1/2}(z_p^{n})+\Delta F_{j+1/2}(y_p^{n})\Delta
  F_{k+1/2}(z_p^{n+1})\}=-\frac14\frac{\Delta x}{\Delta t}\times
$$
$$
\times
\sum_{i^\prime\le i}{\Delta
  F^{(-)}_{i^\prime-1/2}(x_p)}
[\Delta F^{(+)}_{j+1/2}(y_p)\Delta F^{(+)}_{k+1/2}(z_p)+
\frac1{3}\Delta F^{(-)}_{j+1/2}(y_p)\Delta F^{(-)}_{k+1/2}(z_p)],
$$
$$
\xi^{(y)\ n+1/2}_{i+1/2,j,k+1/2}({\bf x}^{n}_p,{\bf x}^{n+1}_p)=-
\frac14\frac{\Delta y}{\Delta t}\sum_{j^\prime\le j}{\Delta
  F^{(-)}_{j^\prime-1/2}(y_p)}\times
$$
$$
\times[\Delta F^{(+)}_{i+1/2}(x_p)\Delta F^{(+)}_{k+1/2}(z_p)+
\frac1{3}\Delta F^{(-)}_{i+1/2}(x_p)\Delta F^{(-)}_{k+1/2}(z_p)],
$$
$$
\xi^{(z)\ n+1/2}_{i+1/2,j+1/2,k}({\bf x}^{n}_p,{\bf x}^{n+1}_p)=-
\frac14\frac{\Delta z}{\Delta t}\sum_{k^\prime\le k}{\Delta
  F^{(-)}_{k^\prime-1/2}(z_p)}\times
$$
\begin{equation}\label{current}
\times[\Delta F^{(+)}_{i+1/2}(x_p)\Delta F^{(+)}_{j+1/2}(y_p)+
\frac1{3}\Delta F^{(-)}_{i+1/2}(x_p)\Delta F^{(-)}_{j+1/2}(y_p)].
\end{equation}
Here
$
\Delta F^{(\pm)}_{j+1/2}(y_p)=\Delta F_{j+1/2}(y^{n+1}_p)\pm\Delta
F_{j+1/2}(y^{n}_p), ...$. 
A recursive formula,  
$
F_i(x^{n+1}_p)-F_i(x^{n}_p)=F_{i-1}(x^{n+1}_p)-F_{i-1}(x^{n}_p)+
\Delta F^{(-)}_{i-1/2}(x_p)$, allows us to calculate
$F_i(x^{n+1}_p)-F_i(x^{n}_p)=\sum_{i^\prime\le i}{\Delta
  F^{(-)}_{i^\prime-1/2}(x_p)}$. 
The scheme as in Eq.(\ref{current}) is not new and was
obtained from different considerations and in a different form in
\cite{esirkepov} (see also discussion in \cite{umeda,barth}).

Note a useful property of the interpolation coefficients. Summing up 
the contributions to the electric current at different faces and using 
Eqs.(\ref{eq:sum1},\ref{eq:sumcurrent}) we
obtain the following {\it exact} relationship:
$$
\sum_{i,j,k}{\xi^{(x)\ n+1/2}_{i,j+1/2,k+1/2}({\bf x}^{n}_p,{\bf
    x}^{n+1}_p)}=u_p^{(x)\ n+1/2},
$$
and the analogous relationships for $y$ and $z$ components. These
relationships for the interpolation coefficients for the electric
current are similar to those for the electric and magnetic fields 
(see \ref{one}). 
\section{Interpolation procedure for the electric and magnetic fields.} 
To interpolate the charge density we had to integrate the form-factor 
over the cell volume. 
Here we show that to 
interpolate the electric and magnetic fields the form-factor should be 
integrated over the cell faces and edges correspondingly.
\subsection{Energy integral}
Consider the energy, ${\cal E}_p$, of the system consisting of the 
electric field and a single charged particle. Both the particle energy and the electric
current contributing to the electric field energy change are additive
by particles, hence, so is the error in the energy
conservation and thus, we can calculate it for a single particle. For simplicity, we assume zero magnetic field
because this field does not affect the particle energy and any 
change in the magnetic field energy is balanced with that for the 
electric field. At time instant, $t=n$, the energy, ${\cal E}^{n}_p$,
equals:
$$
m_pc\sqrt{({\bf w}^{n-1/2}_p)^2+c^2}+\frac V{8\pi}\left(\sum_{i,j,k}{(E^{(x)\ n}_{i,j+1/2,k+1/2})^2}+...\right)+
  \frac{q_p\Delta t}{2}({\bf u}^{n-1/2}_p\cdot{\bf E}({\bf x}^{n}_p)),
$$
the last term is to advance the particle energy through a half time
step. 
The change in the particle energy with the use of Eq.(\ref{p2}), can be
approximated as: 
$$
\sqrt{({\bf w}^{n+1/2}_p)^2+c^2}-\sqrt{({\bf w}^{n-1/2}_p)^2+c^2}\approx
\frac{q_p\Delta t}{2m_pc}\left({\bf E}({\bf x}^{n}_p)\cdot({\bf u}^{n-1/2}_p+{\bf u}^{n+1/2}_p)\right)
,
$$
with the error, $O((\Delta t)^3)$. 
Neglecting this error, we derive the change in the total energy, using Eqs.(\ref{xi},\ref{escheme}): 
$$
{\cal E}_p^{n+1}-{\cal E}_p^{n}=
\frac V{8\pi}\left(\sum_{i,j,k}{[(E^{(x)\ n+1}_{{\rm
        x\ face}})^2-(E^{(x)\ n}_{{\rm x\ face}})^2]}+...\right)+
$$
$$
+\frac{q_p\Delta t}{2}\left(({\bf E}({\bf x}^{n}_p)+{\bf E}({\bf
  x}^{n+1}_p))\cdot{\bf u}^{n+1/2}_p\right)=
$$
$$
\frac{q_p\Delta t}{2}
\left[
({\bf E}({\bf x}^{n}_p)+{\bf E}({\bf
  x}^{n+1}_p))\cdot{\bf
    u}^{n+1/2}_p
-
\sum_{i,j,k}{
\xi^{(x) n+1/2}_{{\rm x\ face}}
[E^{(x)\ n+1}_{{\rm
        x\ face}}+E^{(x)\ n}_{{\rm x\ face}}]}-...\right]=
$$
$$
=\sum_{i,j,k}{\left(\Delta_1{\cal E}^{(x)\ n+1/2}_{i,j+1/2,k+1/2}+
\Delta_3{\cal E}^{(x)\ n+1/2}_{i,j+1/2,k+1/2}\right)+...},
$$
where:
$$
\Delta_1{\cal E}^{(x)\ n+1/2}_{i,j+1/2,k+1/2}=\Delta_1{\cal E}^{(x)\ n+1/2}_{{\rm x\ face}}=\frac{q_p\Delta t}4
(E^{(x)\ n}_{{\rm x\ face}}+E^{(x)\ n+1}_{{\rm x\ face}})\times
$$
$$
\times
\left[
u^{(x)\ n+1/2}_p
\left(
\alpha^{(x)}_{{\rm x\ face}}({\bf x}^{n}_p)+
\alpha^{(x)}_{{\rm x\ face}}({\bf x}^{n+1}_p)\right)
-2\xi^{(x)\ n+1/2}_{{\rm x\ face}}
\right],...,
$$
$$
\Delta_3{\cal E}^{(x)\ n+1/2}_{{\rm x\ face}}=\frac{\pi\left(q_p\Delta t\right)^2}V
\xi^{(x)\ n+1/2}_{{\rm x\ face}}\times
$$
$$
\times u^{(x)\ n+1/2}_p\left(\alpha^{(x)}_{{\rm x\ face}}({\bf x}^{n}_p)-
\alpha^{(x)}_{{\rm x\ face}}({\bf x}^{n+1}_p)\right),...,
$$
``x face'' stands for $i,j+1/2,k+1/2$. If 
the interpolation weights for the electric field match those for the
current in the way as follows:
\begin{equation}\label{match}
u^{(x)\ n+1/2}_p
[
\alpha^{(x)}_{{\rm x\ face}}({\bf x}^{n}_p)+
\alpha^{(x)}_{{\rm x\ face}}({\bf x}^{n+1}_p)]
=2\xi^{(x)\ n+1/2}_{{\rm x\ face}}({\bf x}^{n}_p,{\bf x}^{n+1}_p),...
\end{equation}
then $\Delta_1{\cal E}=0$ and 
the energy defect is 
small: $\Delta_3{\cal E}\sim(\Delta t)^3$. 

The problem is that, in contrast with an ordinary PIC scheme, {\it within the ChCPIC scheme} one can hardly
satisfy (\ref{match}) exactly. 
Eq.(\ref{match}) can be obtained as 
the trapezoidal {\it estimate} 
for the integrals in Eqs.(\ref{Jx}-\ref{Jz}), if the interpolation
weights, $\alpha$, are chosen
as follows: 
$$
\alpha^{(x)}_{i,j+1/2,k+1/2}({\bf x}^{n}_p)=f_i(x^{n}_p) 
\Delta F_{j+1/2}(y^{n}_p)\Delta F_{k+1/2}(z^{n}_p),
$$
$$
\alpha^{(y)}_{i+1/2,j,k+1/2}({\bf x}^{n}_p)= 
\Delta F_{i+1/2}(x^{n}_p)f_j(y^{n}_p)\Delta F_{k+1/2}(z^{n}_p),
$$
\begin{equation}\label{alphax}
\alpha^{(z)}_{i+1/2,j+1/2,k}({\bf x}^{n}_p)= 
\Delta F_{i+1/2}(x^{n}_p)\Delta F_{j+1/2}(y^{n}_p)f_k(z^{n}_p).
\end{equation}
The accuracy of Eq.(\ref{match})
and the energy defect while using
Eqs.(\ref{Jx}-\ref{current},\ref{alphax}) are controlled by the choice
of the form-factor order. 
Using the 
 estimate:
$$
\frac{\Delta
  x}{\Delta t}[F_i(x_p^n)-F_i(x_p^{n+1})]=\frac{\Delta
  x}{\Delta t}\int_{i-x_p^{n+1}}^{i-x_p^n}{f(x-x_p)d(x-x_p)}=
$$
$$
=u^{(x)\ n+1/2}_p\left\{\frac12\left[f_i(x_p^n)+f_i(x_p^{n+1})\right]-u^{(x)\ n+1/2}_p\frac{\Delta
  t}{\Delta x}\int_{-1/2}^{1/2}{\frac{df(x-x_p)}{dx}g\,dg}\right\},
$$
where we substituted
$$x-x_p=i-(x_p^n+x_p^{n+1})/2+g(x_p^{n+1}-x_p^n)=i-(x_p^n+x_p^{n+1})/2+gu^{(x)\ n+1/2}_p\Delta
t/\Delta x,$$
and noting that for $l=0,1,2$ correspondingly the form-factor, $f(x-x_p)$, its first or second
derivative are bounded, 
we can evaluate the accuracy of the
energy conservation using the above estimate: 
$\Delta_1{\cal E}\sim O((\Delta t)^{l+1})$. If $l=0$ or, alternatively,
if Eq.(\ref{match}) is not 
fulfilled at all, the
energy does not conserve in the ChCPIC scheme: $\Delta_1{\cal E}\sim
O(\Delta t)$. 
We conclude that both the use of higher-order form-factor ($l\ge1$) and
the interpolation following Eq.(\ref{alphax}) are desirable. 
\subsection{Generalized momentum conservation}
The governing equations (\ref{1}-\ref{vectorpot}) conserve the  projection of the 
generalized particle momentum, $m_p{\bf w}_p+q_p{\bf A}/c$, on the given direction ${\bf g}$,
if the electromagnetic field is constant along this direction, i.e. $({\bf
  g}\cdot\nabla){\bf A}=0$ (see \cite{landau80}). Indeed, 
\begin{equation}\label{momentumcons}
\frac d{dt}(m_p{\bf w}_p+\frac{q_p}c{\bf A})=\frac{q_p}c\left([{\bf u}_p\times
[\nabla\times{\bf A}]]+
({\bf u}_p\cdot\nabla){\bf A}\right)=\frac{q_p}c\nabla\left({\bf u}_p\cdot{\bf A}\right),
\end{equation}
and ${\bf g}\cdot\frac d{dt}(m_p{\bf w}_p+q_p{\bf A}/c)=0$ as long as 
$({\bf g}\cdot\nabla){\bf A}=0$. 
To verify the generalized momentum conservation within the ChCPIC scheme,
the latter should be formulated in terms of the vector potential. The grid
functions $A^{(x)}_{i,j+1/2,k+1/2}$, $A^{(y)}_{i+1/2,j,k+1/2}$, 
$A^{(z)}_{i+1/2,j+1/2,k}$ are introduced at the same points as 
$E^{(x)}_{i,j+1/2,k+1/2}$, $E^{(y)}_{i+1/2,j,k+1/2}$, 
$E^{(z)}_{i+1/2,j+1/2,k}$ 
 and the time derivative of the vector potential grid function may be
 expressed in terms of the electric field:
$$
A^{(x)\ n+1/2}_{i,j+1/2,k+1/2}=A^{(x)\ n-1/2}_{i,j+1/2,k+1/2}-c\Delta t E^{(x)\ n}_{i,j+1/2,k+1/2},
$$
$$
A^{(y)\ n+1/2}_{i+1/2,j,k+1/2}=A^{(y)\ n-1/2}_{i+1/2,j,k+1/2}-c\Delta t E^{(y)\ n}_{i+1/2,j,k+1/2},
$$
\begin{equation}\label{dadt}
A^{(z)\ n+1/2}_{i+1/2,j+1/2,k}=A^{(z)\ n-1/2}_{i+1/2,j+1/2,k}-c\Delta t E^{(z)\ n}_{i+1/2,j+1/2,k},
\end{equation}
in accordance with the first of
Eqs.(\ref{vectorpot}). 
The electric field and the vector 
potential at ${\bf x}={\bf x}_p$ are also linked via this equation,
hence, the vector potential in a particle location, ${\bf A}(t,{\bf x}_p)$ should be
interpolated with the same weights, $\alpha$, as we use for
interpolating the electric field:
\begin{equation}\label{A}
A^{(x)}(t,{\bf x}_p)=
\sum_{i,j,k}
{A^{(x)\ t}_{i,j+1/2,k+1/2}
\alpha^{(x)}_{i,j+1/2,k+1/2}({\bf x}_p)},...
\end{equation}

On the other hand, 
 the magnetic field acting on the particle can be interpolated via the
 grid function of the magnetic field,
 $B^{(x)}=\sum_{i,j,k}{\beta^{(x)}_{i+1/2,j,k}B^{(x)}_{i+1/2,j,k}}$. In
 turn, the
 latter grid function can be expressed in terms of that for the vector potential,
 as the discretization of the second of Eqs.(\ref{vectorpot}), on the
 staggered grid:
$$
B^{(x)}_{i+1/2,j,k}=\frac1{\Delta
  y}\left(A^{(z)}_{i+1/2,j+1/2,k}-A^{(z)}_{i+1/2,j-1/2,k}\right)-
$$
$$
-\frac1{\Delta z}\left(A^{(y)}_{i+1/2,j,k+1/2}-A^{(y)}_{i+1/2,j,k-1/2}\right),...,
$$
 so that:
\begin{eqnarray}
\nonumber B^{(x)}(t,{\bf x}_p)=\frac{1}{\Delta y}\sum_{i,j,k}
{(\beta^{(x)}_{i+1/2,j,k}-\beta^{(x)}_{i+1/2,j+1,k})A^{(z)\ t}_{i+1/2,j+1/2,k}}-\\ 
-\frac{1}{\Delta z}\sum_{i,j,k}
{(\beta^{(x)}_{i+1/2,j,k}-\beta^{(x)}_{i+1/2,j,k+1})
A^{(y)\ t}_{i+1/2,j,k+1/2}},...
\label{viaA}
\end{eqnarray}
Now, we introduce the generalized momentum, 
$$
{\bf P}^{n-1/2}=
m_p{\bf w}^{n-1/2}_p+\frac{q_p}c
{\bf A}(n-1/2,{\bf x}^{n}_p-{\bf w}^{n-1/2}\frac{\Delta t}2).
$$ 
Evaluating the difference ${\bf P}^{n+1/2}-{\bf P}^{n-1/2}$, we see that the 
transformation as in Eq.(\ref{momentumcons}), which allows the
generalized momentum conservation, is
possible, if {\it the differential equation, ${\bf
    B}=\nabla\times{\bf A}$, is exactly fulfilled with the interpolated
  values of the magnetic field, and vector potential}, i.e.
\begin{equation}\label{eq:A2B}
{\bf B}({\bf x}_p)=[\nabla_p\times {\bf A}(t,{\bf x}_p)],
\end{equation} 
where 
$\nabla_p=(\frac1{\Delta x}\frac\partial{\partial x_p},..)$. Applying
the operator, $\nabla_p\times$ to Eq.(\ref{A}) and comparing the result
with Eq.(\ref{viaA}) we find that Eq.(\ref{eq:A2B}) is fulfilled if the
following set of equations holds:
$$
\beta^{(x)}_{i+1/2,j,k}-\beta^{(x)}_{i+1/2,j+1,k}=
\frac{\partial\alpha^{(z)}_{i+1/2,j+1/2,k}}{\partial y_p},
$$
$$ 
\beta^{(x)}_{i+1/2,j,k}-\beta^{(x)}_{i+1/2,j,k+1}=
\frac{\partial\alpha^{(y)}_{i+1/2,j,k+1/2}}{\partial z_p},
$$ 
etc. These equations as well as  Eq.(\ref{alphax}) dictate the following choice for the
magnetic field weights:
$$
\beta^{(x)}_{i+1/2,j,k}=\Delta
F_{i+1/2}(x^{n}_p)f_j(y^{n}_p)f_k(z^{n}_p),\quad 
\beta^{(y)}_{i,j+1/2,k}=f_i(x^{n}_p)\Delta F_{j+1/2}(y^{n}_p)f_k(z^{n}_p),
$$
\begin{equation}\label{betax}
\beta^{(z)}_{i,j,k+1/2}=f_i(x^{n}_p)f_j(y^{n}_p)\Delta F_{k+1/2}(z^{n}_p).
\end{equation}
\subsection{Electrostatic limit and a link to the finite element approach}
To discuss the relationship with the important concepts of the momentum 
conservation and self-force (see detail in\cite{birdsall85}) we now consider the electrostatic limit in which the fields acting on a given macroparticle can 
be assumed steady-state. Within this approximation, the electric field may be 
expressed in terms of the scalar electric potential, $\Phi$:
\begin{equation}\label{eq:egradphi}{\bf E}=-\nabla\Phi\end{equation}
The grid function for the scalar potential is cell-centered, $\Phi_{i+1/2,j+1/2,k+1/2}$. The relationships for the face-centered electric field components are as follows:
\begin{equation}\label{eq:egradphidis}E^{(x)}_{i,j+1/2,k+1/2} =\frac1{\Delta x}\left(\Phi_{i-1/2,j+1/2,k+1/2}-\Phi_{i+1/2,j+1/2,k+1/2}\right),...\end{equation}
The Poisson equation should be fulfilled in the following form:
\begin{eqnarray}\label{eq:poiphi}\frac{-\Phi_{i-1/2,j+1/2,k+1/2}+2\Phi_{i+1/2,j+1/2,k+1/2}-
\Phi_{i+3/2,j+1/2,k+1/2}}{(\Delta x)^2}+...=\nonumber\\
=4\pi\rho_{i+1/2,j+1/2,k+1/2}.
\end{eqnarray}
With this observations, the contribution from the electric field to the energy 
integral may be transformed using Eqs.(\ref{eq:egradphidis},\ref{eq:poiphi}) 
as follows:
$$
\frac V{8\pi}\left(\sum_{i,j,k}{(E^{(x)\ n}_{i,j+1/2,k+1/2})^2}+...\right)=\frac{V}2\sum_{i,j,k}{\rho_{i+1/2,j+1/2,k+1/2}\Phi_{i+1/2,j+1/2,k+1/2}}.
$$
The key point is that in the last sum the charge density can be expressed in terms of the contributions from the given particles, then resulting in the field 
energy transformation to the sum over all charged particles:
$$
\frac{V}2\sum_{i,j,k}{\rho_{i+1/2,j+1/2,k+1/2}\Phi_{i+1/2,j+1/2,k+1/2}}=\frac12\sum_p{q_p\Phi({\bf x}_p}),
$$ 
where the static potential at the particle location is introduced as follows:
\begin{equation}
\Phi({\bf x}_p)=\sum_{i,j,k}{\Delta F_{i+1/2}(x_p)\Delta F_{j+1/2}(y_p)\Delta F_{k+1/2}(z_p)\Phi_{i+1/2,j+1/2,k+1/2}}.
\end{equation}
If the energy is conserved, then the energy integral may play the role of the 
Hamiltonian function, so that the electric force acting on a particle can be 
expressed in terms of partial derivatives of the Hamiltonian function over 
particle coordinates. As the result, the electric field acting on the 
particle should obey Eq.(\ref{eq:egradphi}):
$$
{\bf E}({\bf x}_p)=-\nabla_p\Phi({\bf x}_p).
$$
But this is achieved with the choice of Eq.(\ref{alphax}) for the weight 
coefficients, to interpolate the electric field acting on a particle! Indeed,
$$
-\frac1{\Delta x}\frac\partial{\partial x_p}\sum_{i,j,k}{\Delta F_{i+1/2}(x_p)\Delta F_{j+1/2}(y_p)\Delta F_{k+1/2}(z_p)\Phi_{i+1/2,j+1/2,k+1/2}}=$$ $$=
\frac1{\Delta x}\sum_{i,j,k}{(f_{i+1}(x_p)-f_{i}(x_p))\Delta F_{j+1/2}(y_p)\Delta F_{k+1/2}(z_p)\Phi_{i+1/2,j+1/2,k+1/2}}=$$ $$=
\sum_{i,j,k}{\alpha^{(x)}_{i,j+1/2,k+1/2}E^{(x)}_{i,j+1/2,k+1/2}}=E^{(x)}_p({\bf x}_p),...
$$
Here, the conclusion of Langdon is reproduced, that the ``energy-conserving'' 
plasma simulations models require the contribution from the potential 
electric field to the force acting on a particle to be a gradient of the grid 
potential (see \cite{Langdon}). From here yet another way to arrive at the 
suggested scheme for interpolating the electric and magnetic force may be 
recalled, specifically, the finite elements. The nodal amplitudes for fields 
and the form-factors are equivalent to local polynomial representation for the 
field using finite elements. The introduction of incomplete polynomial face 
(div-conforming) and edge (curl-conforming) elements is generally attributed 
to N\'ed\'elec \cite{Nedelec}. For the ``energy-conserving'' scheme presented 
by Lewis in \cite{Lewis} (see also \cite{Hockney}) the use of such finite 
elements allows maintaining some of the Maxwell equations as the exact 
relationships between the finite elements for the fields. In the 
application to arbitrary curvilinear grids made in \cite{eastwood2} this 
approach resulted in the same form-factors as in 
 Eqs.(\ref{alphax},\ref{betax}), the form-factor order being $l=1$. 

Nevertheless, the new derivation of Eqs.(\ref{alphax},\ref{betax}) for newer 
ChCPIC deserves the attention, moreover that the formulation of the 
conservation laws for such schemes is different from the formulation of 
\cite{Lewis}. We also emphasized that for the form-factor order of $l=1$ 
there is no point in applying the interpolation scheme of 
Eqs.(\ref{alphax},\ref{betax}), as long as the energy (nor the generalized 
momentum) is not conserved anyway with this form-factor.     
\subsection{1D scheme and self-force}
To illustrate the entire scheme, consider one-dimensional (1D) motion of a 
single particle in the given steady-state fields depending on $x$-coordinate 
only. Assume zero transverse components of the electric field as well as a zero 
longitudinal component of the magnetic field. Eqs.(\ref{alphax},\ref{betax}) 
in 1D read:
$$E^{(x)}(x_p)=\sum_i{f_i(x_p)E^{(x)}_i}=-\frac1{\Delta x}\frac{\partial}{\partial x_p} \Phi(x_p),$$ $${\bf B}^{(\perp)}(x_p)=\sum_i{f_i(x_p){\bf B}_i^{(\perp)}}=\frac{1}{\Delta x}\frac\partial{\partial x_p}{\bf n}_x\times {\bf A}^{(\perp)}(x_p),$$
where the subscript $\perp$ denotes the tranverse components of the vector 
potential and magneitc field and ${\bf n}_x$ is the vector along $x$-coordinate. In the limit of an infinitely small timestep and in negecting for a while 
the particle charge effect on the fields, we find that three integrals fully describe 
the mactroparticle motion:
$$
{\bf p}^{(\perp)}+\frac{q_p}c{\bf A}^{(\perp)}(x_p)=const,\qquad{\cal E}+q_p\Phi(x_p)=const,
$$
where, again, the continuous potentials at the particle position are expressed 
in terms of cell-centered grid function values. Within this approximation, 
however, the motion description is perfectly accurate, as long as the 
conservation laws are fulfilled which fully control the particle motion. 

However, as noticed in \cite{birdsall85},\cite{Langdon}, the energy 
conservaion property of the numerical scheme results in the side effect in a 
form of a  self-force. The conserved energy includes the contribution from the 
particle's own field energy, which is weakly dependent on the particle location 
relative to the grid nodes. As a result the weak force appears tending 
to pull the particle from cell centers towards faces.

Now one can calculate the self-force effect on the 1D macroparticle motion. The current 
through $x$-faces, 
$-\frac{\Delta X}{\Delta t}\frac{q_p}V\left(F_i(x_p^{n+1})-F_i(x_p^n)\right)$, can 
be integrated over time, giving the 
electric 
field: $E^{(x)}_i=\frac{4\pi q_p\Delta x}{V}\left(F_i(x_p)-\frac12\right)$. 
Now we apply the weight coefficients as in Eq.(\ref{alphax}) to obtain the 
self-force acting on the macroparticle in the form of the gradient of the 
particle's own field energy:
$$q_p\sum_i{f_i(x_p)E^{(x)}}=-\frac{q_p}{\Delta x}\frac{\partial \Phi^{(own)}(x_p)}{\partial x_p},$$
$$
 q_p\Phi^{(own)}(x_p)=\frac{4\pi q_p^2\Delta^2 x}V\sum_i
{\frac{(F_i(x_p)-1/2)^2}2}.
$$
With this accounting, the energy conservation law provides the 
distorted result: 
${\cal E}+q_p\left(\Phi(x_p)+\Phi^{(own)}(x_p)\right)=const$. Upon calculating the transverse magnetic field and vector potential induced by the particle 
{\it slow} motion, one can find that the transverse particle momentum is 
distorted accordingly, so that the ratio, which is the transverse velocity, 
is not modified with the particle's own field. So, {\it in 1D non-relativistic motion  
the self-force does not affect the transverse velocity}. 

To evaluate the self-force effect on the longitudinal velocity, note that the  
dimensional factor in the formula for $q_p\Phi^{(own)}(x_p)$ as related to 
temperature equals $\Delta^2 x/(Nr_D^2)$, where $r_D$ is the Debye length and
$N$ is the number of macroparticles per cell. Usually (but not in the 
numerical tests discussed below) this multiplier is small, which reduces the 
effect of the self-force comparing to the macroparticle interaction with the 
regular electromagnetic field and with the neighboring particles. The 
dimensionless factor was evaluated in \cite{birdsall85} for the lowest-order form-factor and the 
discontinuous character of the self-force was noticed for this case, 
resulting in potential problems. However, for the choice 
of a higher order form-factor (which is called ``form-factor of the 
order $l=1\frac12$'' below), the field energy is not 
only a smooth function of $x_p$, but its variance is very small:
$$
\sum_i
{\frac{(F_i(x_p)-1/2)^2}2}=\frac18+\frac14(x_p-J)^2(J+1-x_p)^2,\qquad 
J\le x_p\le J+1.         
$$
The variance 
from minimum to maximum is 1/64. We see that the self-force is rather small 
and purely potential and its effect was not revealed in the numerical tests 
discussed below.   
\subsection{Alternating-order form-factor}
\begin{figure}\label{Fig1}
\includegraphics[scale=0.3]{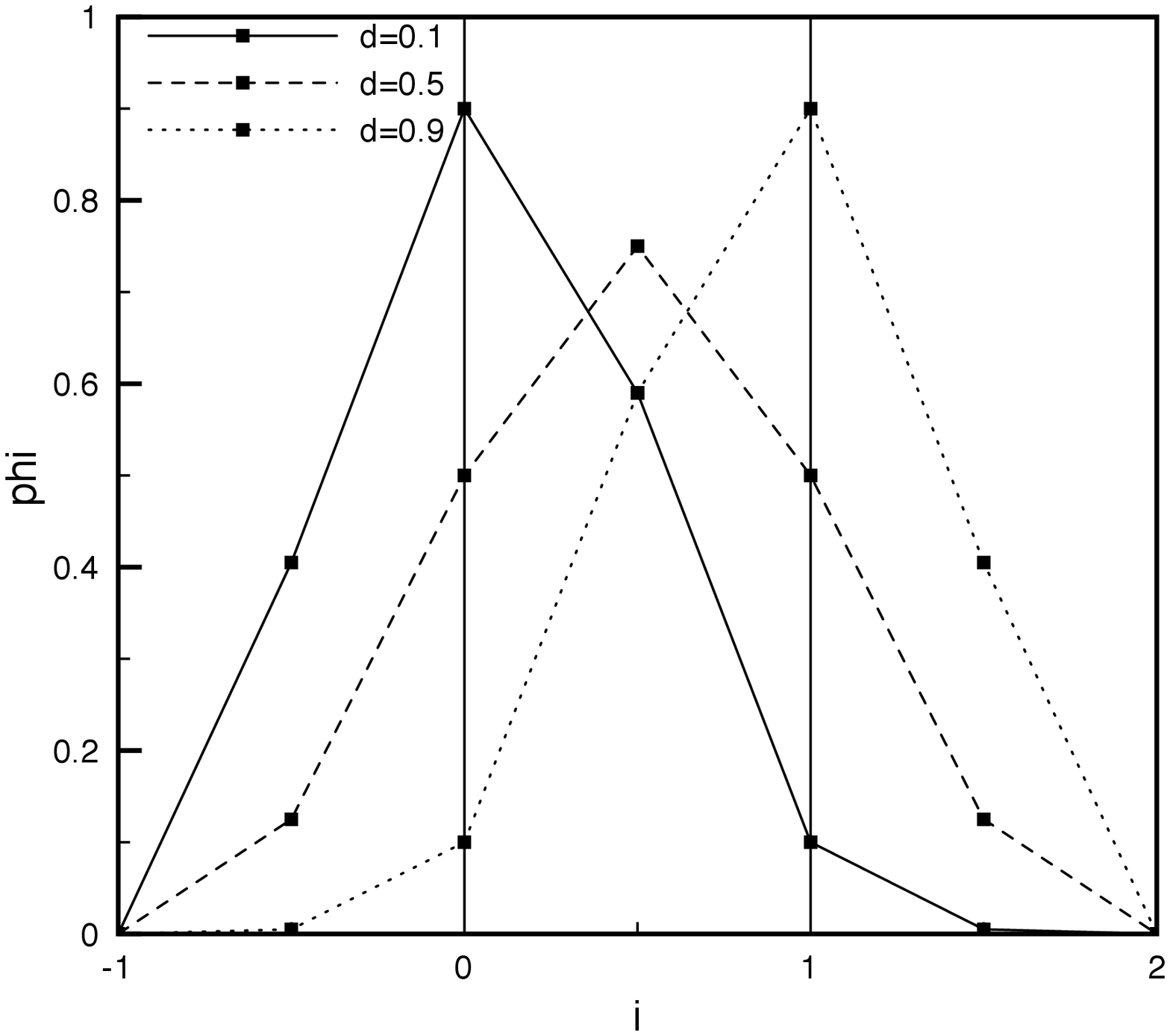}
\includegraphics[scale=0.3]{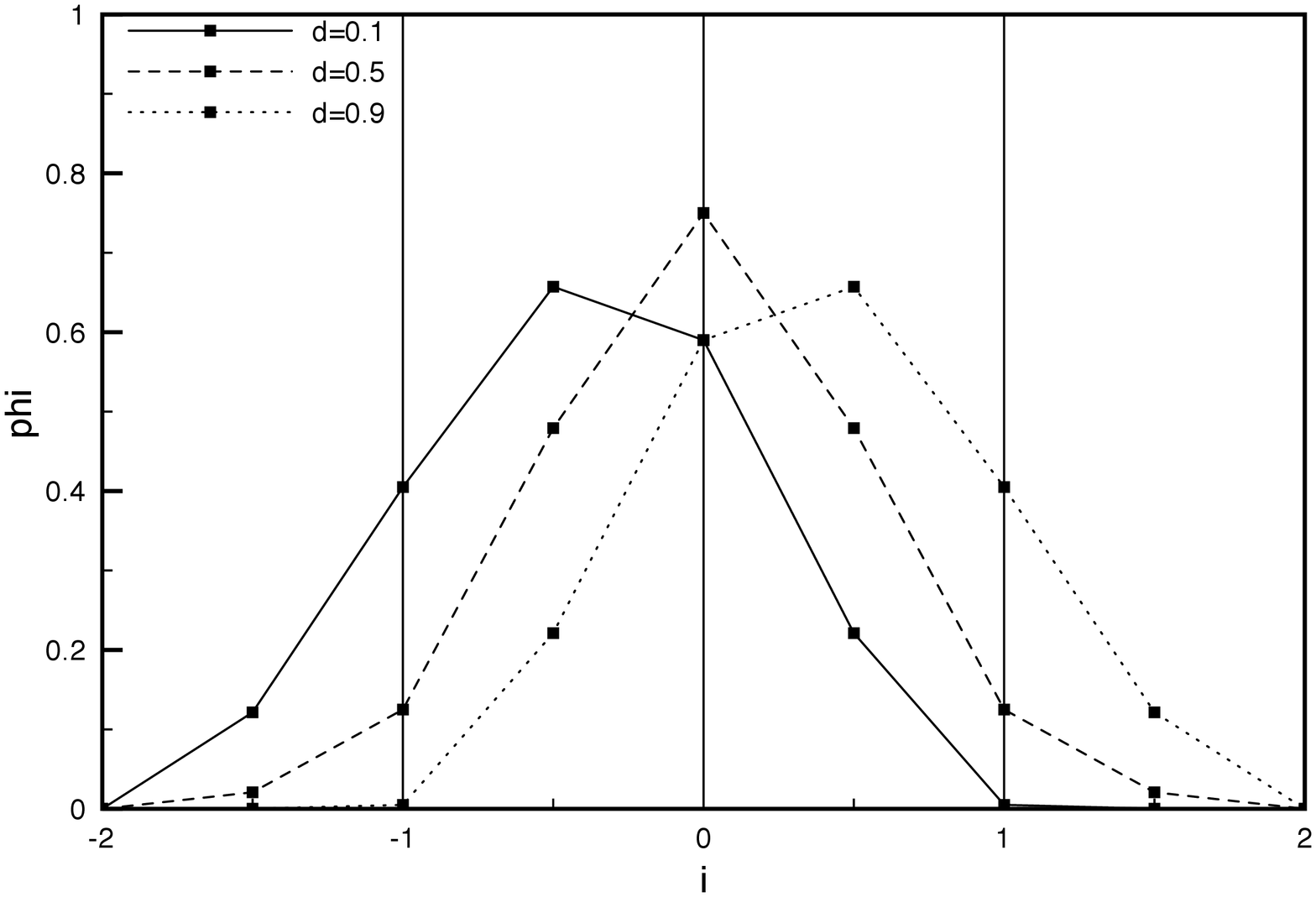}
\caption{Alternating order form-factors (shape functions) of the order
  $1\frac12$ (left panel) and $2\frac12$ (right panel), for different
  values of $x_p$, which are parameterized via $d$. The point values
  are presented for integer values of $x=i$ (marked with the vertical
  lines) and for semi-integer values of $x$.}
\end{figure}
Eqs.(\ref{rho},\ref{alphax},\ref{betax}) may be interpreted 
in terms of the {\it alternating-order} form-factor,
$\varphi(x,x_p)\varphi(y,y_p)\varphi(z,z_p)$, with
the value of this function at the grid point, $x,y,z$, giving the
interpolation weight for the grid function defined at this
point. Although the form-factor is a continuous function, only 
semi-integer and integers $x,y,z$ matter, to which the grid functions
are assigned. 
Thus, we define 
$\varphi(x,x_p)=f_i(x_p)=f^{(l)}(i,x_p)$ at integer $x$, but $\varphi(x,x_p)=\Delta
F_{i+1/2}(x_p)=f^{(l+1)}(i+1/2,x_p)$ at semi-integer $x$,
i.e. we alternate the form-factor order. As long as the
alternating-order form-factor combines the values of functions,
$f^{(l)}$ and $f^{(l+1)}$, we would characterize it by the semi-integer
order, $l_a=l+1/2$ 
  
Finally, we provide the values of the alternating-order form-factor for
$l_a=1\frac12$:
$$
i={\rm int}(x_p),\quad d=x_p-i,\quad f_{i:i+1}(x_p)=(1-d;d),
$$
$$
\Delta F_{i-1/2:i+3/2}(x_p)=\left(\frac{(1-d)^2}2;\frac 34-(\frac
12-d)^2;
\frac{d^2}2\right),
$$
and for $l_a=2\frac12$, which really ensures high accuracy and good energy conservation:
$$
i={\rm int}(x_p+\frac12),\quad d=x_p+\frac12-i,\quad f_{i-1:i+1}=\left(\frac{(1-d)^2}2;\frac 34-(\frac 12-d)^2;\frac{d^2}2\right),
$$
$$
\Delta F_{i-3/2:i+3/2}(x_p)=\left(\frac{(1-d)^3}6;\frac{(2-d)^3}6-\frac{2(1-d)^3}3;
\frac{(1+d)^3}6-\frac{2d^3}3;\frac{d^3}6\right).
$$
\section{Numerical tests}
The tests we present combine some features of the tests for
ChCPIC as developed in \cite{umeda,barth}. The tests are highly demanding: 
the mesh size is as large as the Debye length, the time step is as large as 
a half of $\Delta x/c$, the macroparticle density is as small as two per cell 
in a three-dimensional case, thus increasing the macroparticle charge. Increased 
effects of particle-particle and particle-grid interactions allow for producing a 
noticeable numerical error within a short test runs. 

First, we study the noise in the plasma with comparatively high level of 
thermal fluctuations. We use a three-dimensional 64*64*64 rectangular grid. 
The average number of electron particles per grid cell is $N=2$. The electron
thermal speed is: $v_{Te}=\sqrt{T_e/m_e}=0.05c$, where $T_e$ is the
electron temperature. At the initial time instant $2*64^3$ electron
particles are randomly distributed over the computational domain,
the averaged plasma frequency being equal to $\omega_{pe}=1$. The equal
number of immovable ions are put to the same
locations as the electrons, so that the electromagnetic field is zero
initially. The 
mesh size is $\Delta x=\Delta y=\Delta z = v_{Te}/\omega_{pe}$. With
the time step, $\omega_{pe}\Delta t = 0.025$, the plasma evolution has been
simulated through two plasma wave periods, $n\omega_{pe}\Delta t\approx
4\pi,\ \,n=503$. Fortran 90/95 code is compiled and run with the double 
precision.

\begin{figure}\label{Fig2}
\includegraphics[scale=0.28]{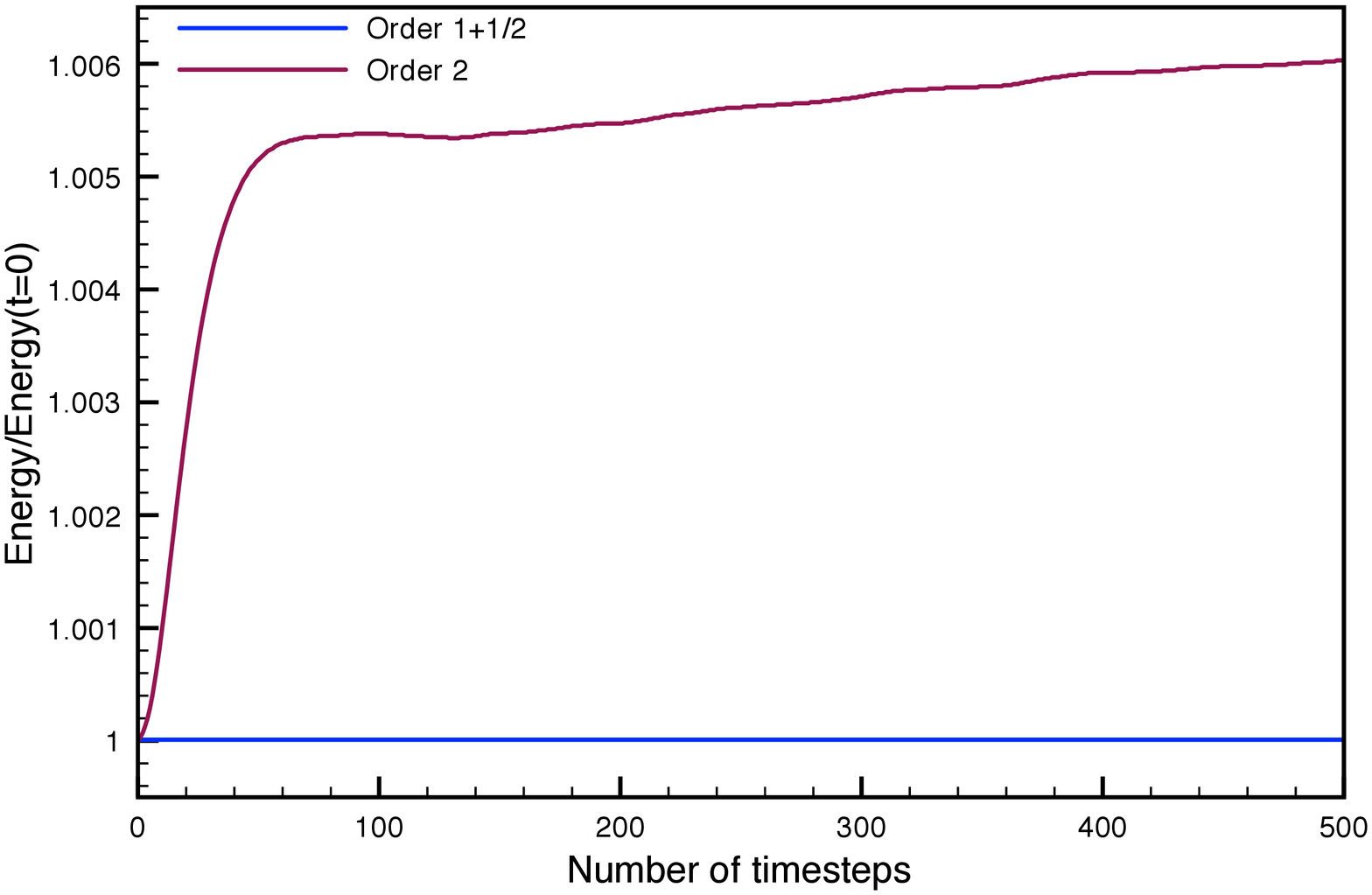}
\includegraphics[scale=0.28]{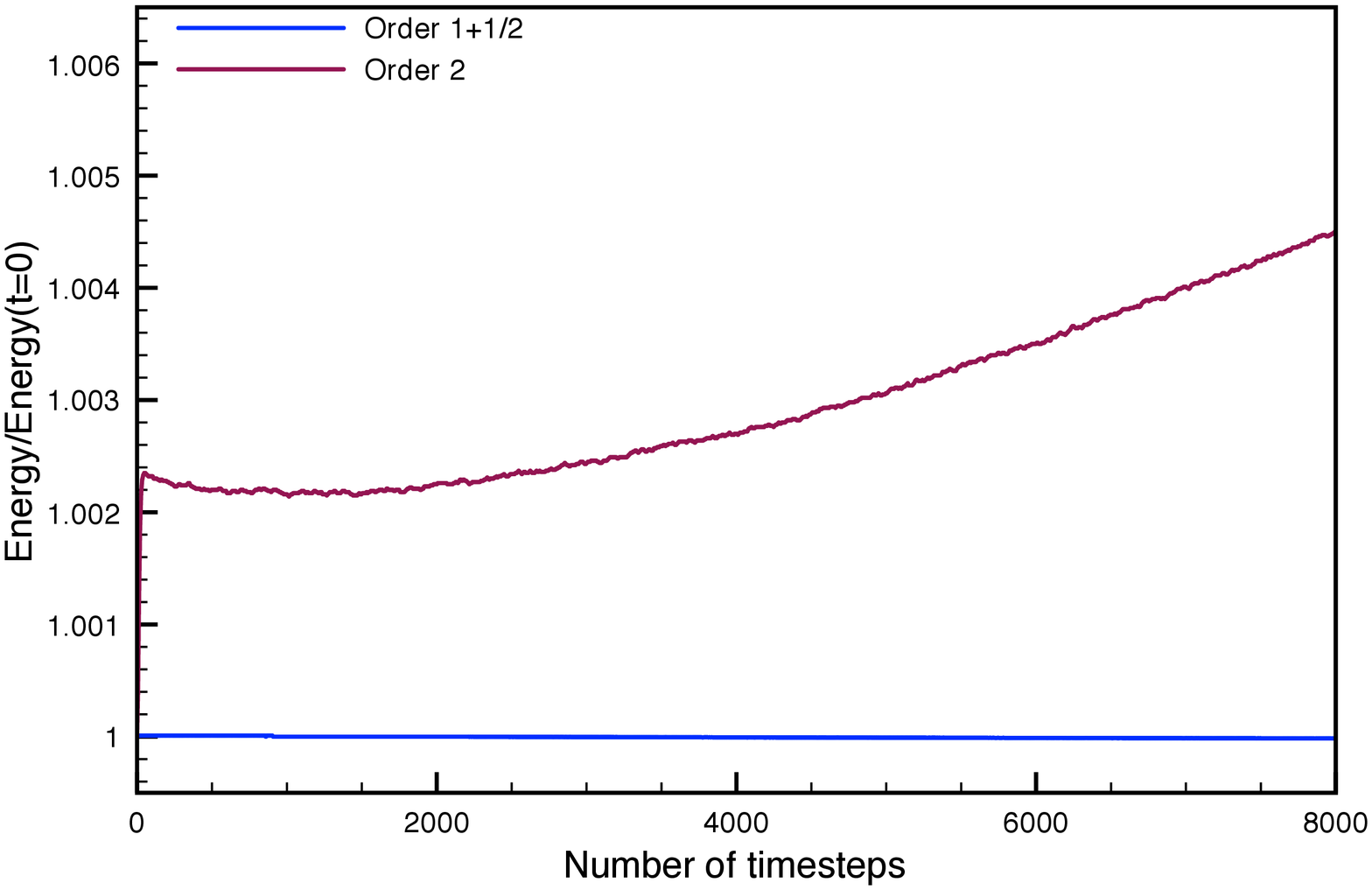}
\caption{Evolution of the total energy
normalized per the initial total energy. Blue curves demonstrate 
perfect energy conservation as achieved with the alternating-order 
form-factor of the order $l_a=1\frac12$. The pink curves show much worse
result obtained with {\it higher} (but not alternating) order
$l=2$. The left panel is for the Maxwellian plasma simulated through
two plasma wave periods. The right panel is for two-stream instability
followed till $\omega_{pe}t=200$ (8000 iterations).}
\end{figure}
In Fig.2 (left panel) we present the evolution of the total energy
normalized per the initial total energy. The blue curve demonstrates an
almost perfect energy conservation as achieved with the alternating-order 
form-factor of the order $l_a=1\frac12$. The pink curve shows, for 
comparison, the result obtained with the ``conventional'' way to interpolate the fields in the particle
location, as used, for example in \cite{barth} as well as in older
works, including the TRISTAN code \cite{buneman2}. Within this approach
the cell-centered form-factor provides the interpolation weight, which is applied to 
the electric field vector as averaged over the faces of the given
cell, which results in the following expression for $\alpha$;
\begin{equation}\label{eq:wrong1}
\alpha_{i,j+1/2,k+1/2}^{(x)}({\bf
  x}_p)=\frac{\sum_\pm{\Delta F_{i\pm1/2}(x_p)}}2\Delta F_{j+1/2}(y_p)\Delta F_{k+1/2}(z_p),...
\end{equation}
and the analogous principle is applied to construct
$\beta$-coefficients, with 
Eq.(\ref{current}) being used to interpolate the electric currents, 
With this approach {\it higher} order $l=2$ is used to
interpolate both the charge density and the electric and magnetic
field.

We clearly see that the {\it increase} in the interpolation
order for the fields, comparing to the alternating-order form-factor results in the
{\it degrade} of an accuracy, as long as the energy defect becomes two
orders of magnitude higher. This numerical result entirely agrees with
our theoretical claims that within ChCPIC scheme the energy conservation
degrades unless the alternating-order form-factor is used.

Note also that the worse quality of the results with the uniform order $l=2$ 
form-factor is accompanied by the degrade in efficiency. The stencil
for $l=2$ form-factor is wider than that for $l_a=1\frac12$, 
therefore, more search/algebra operations are involved in this
case. Thus, the efficiency drops from 
$\approx2.1\cdot10^5$ particles advanced per second of CPU time per
processor, for $l_a=1\frac12$ case, down to $\approx1.5\cdot10^5$
particles/s/processor, for $l=2$ case. 

In the second test taken from \cite{barth} the two-stream instability is
studied. The initial distribution is modified as
follows. First, the magnetic field is applied along the direction of
$x$-axis, such that the cyclotron frequency of electrons in this field equals one:
$\omega_{Be}=q_eB/(m_ec)=1=\omega_{pe}$. To set the unstable distribution function
with two streams, for $50\%$ of the electron particles the directed
stream velocity, $v=2v_{Te}=0.1c$,
along $x$-direction is added to the thermal random velocity. For the other half of the
electron particles the negative of the first stream velocity,
$-v=-2v_{Te}=-0.1c$, is added. The instability evolution is traced
through the time interval, $0\le\omega_{pe}t\le200$. In the right panel of
Fig.2, the check of the total energy conservation, again, demonstrates
the advantage of the alternating-order form-factor (herewith the
contribution from the initial magnetic field is excluded from the total
energy integral).

In Fig.3 we present two test results obtained with the increased
particle charge, so high that $N=1$ particle per cell simulates the
plasma of unity plasma frequency $\omega_{pe}=1$, i.e. the charge is
two times higher compared to the previous tests. Left panel presents the 
Langmuir oscillations. The initial distribution is simular to that for
studying the two-stream instability, but the number of particles (of
two times higher charge) is by a factor of 0.5
less. The unidirectional stream velocity, $v=2v_{Te}=0.1c$, is 
added to all electron particles. The left panel of Fig.3 shows the total 
energy of electrons (upper curve) and the electric field energy, as 
functions of $\omega_{pe}t$. 
The test parameters (such as the low number of particles, resulting in 
comparatively strong electron-ion correlations) are intentionally
chosen to make the errors noticeable, so we can observe gradual
attenuation of the oscillations, as the result of electron-ion
interactions. However, the defect in the {\it total} energy is as small
as $\sim10^{-5}$,
 thanks to using the alternating order from-factor. This defect is negligible
compared to the changes in the partial energy integrals.  
In the right panel, we show the trajectories of particles
gyrating in the magnetic field. The magnetic field is 
increased ten times: $\omega_{Be}=10$, and it is applied along $z$-axis. Two
particles are located initially in the center of the computational domain,
with oppositely directed initial momenta, $w=\pm0.5c$, the particle
charges have opposite signs (electron and positron particles). The
particle trajectories are shown in the right panel of Fig.3 for the
time interval, $0\le\omega_{Be}t\le400$ with the timestep,
$\omega_{Be}\Delta t=0.25$. One can observe a small gradual
decrease in the particle kinetic energy. It reveals itself in the curve
``thickness'', with its inner radius, which is the Larmor radius at the
final particle energy, being somewhat less than the outer radius, which is the Larmor
radius at the initial particle energy. 
   
\begin{figure}
\includegraphics[scale=0.3]{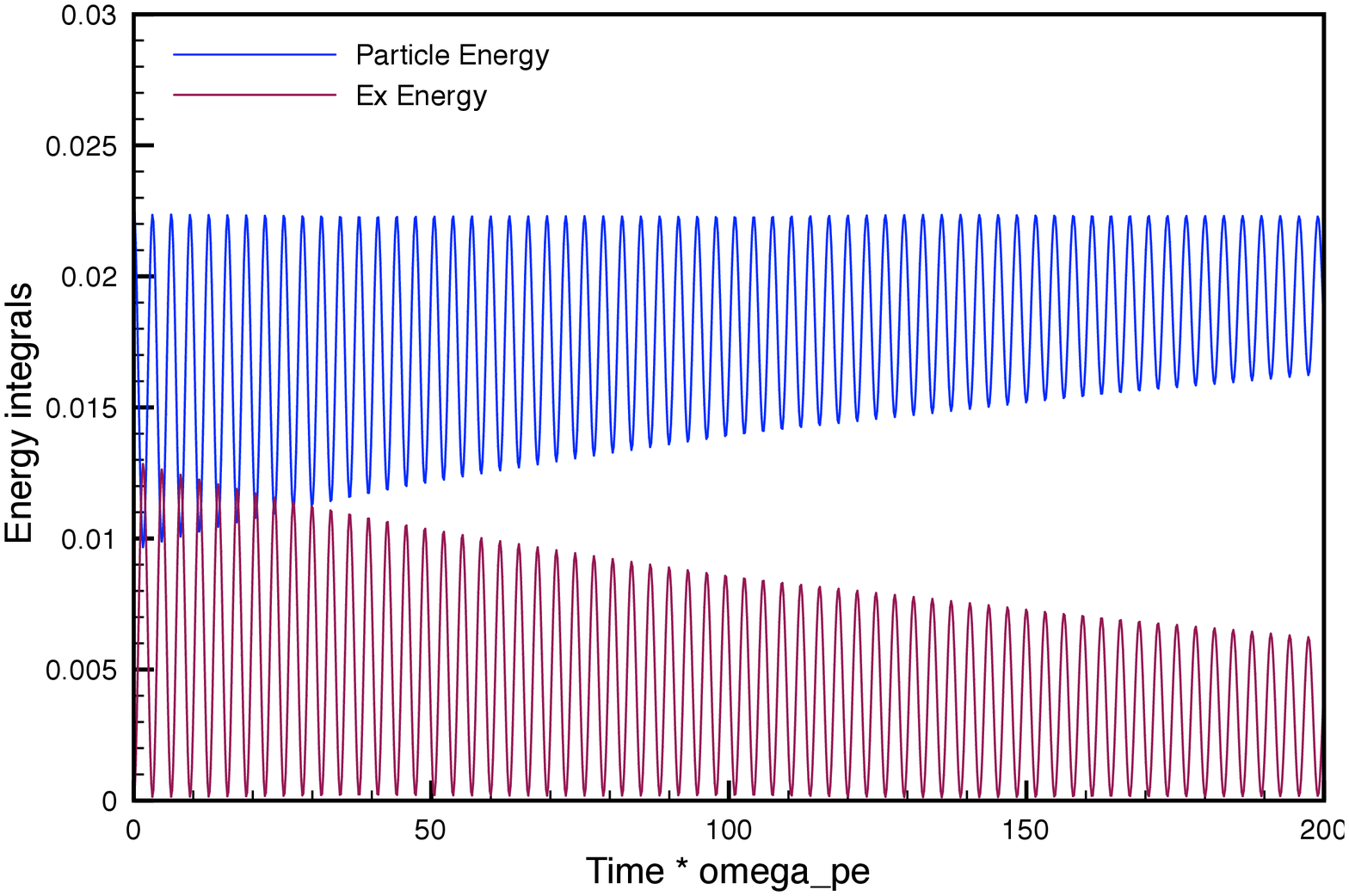}
\includegraphics[scale=0.3]{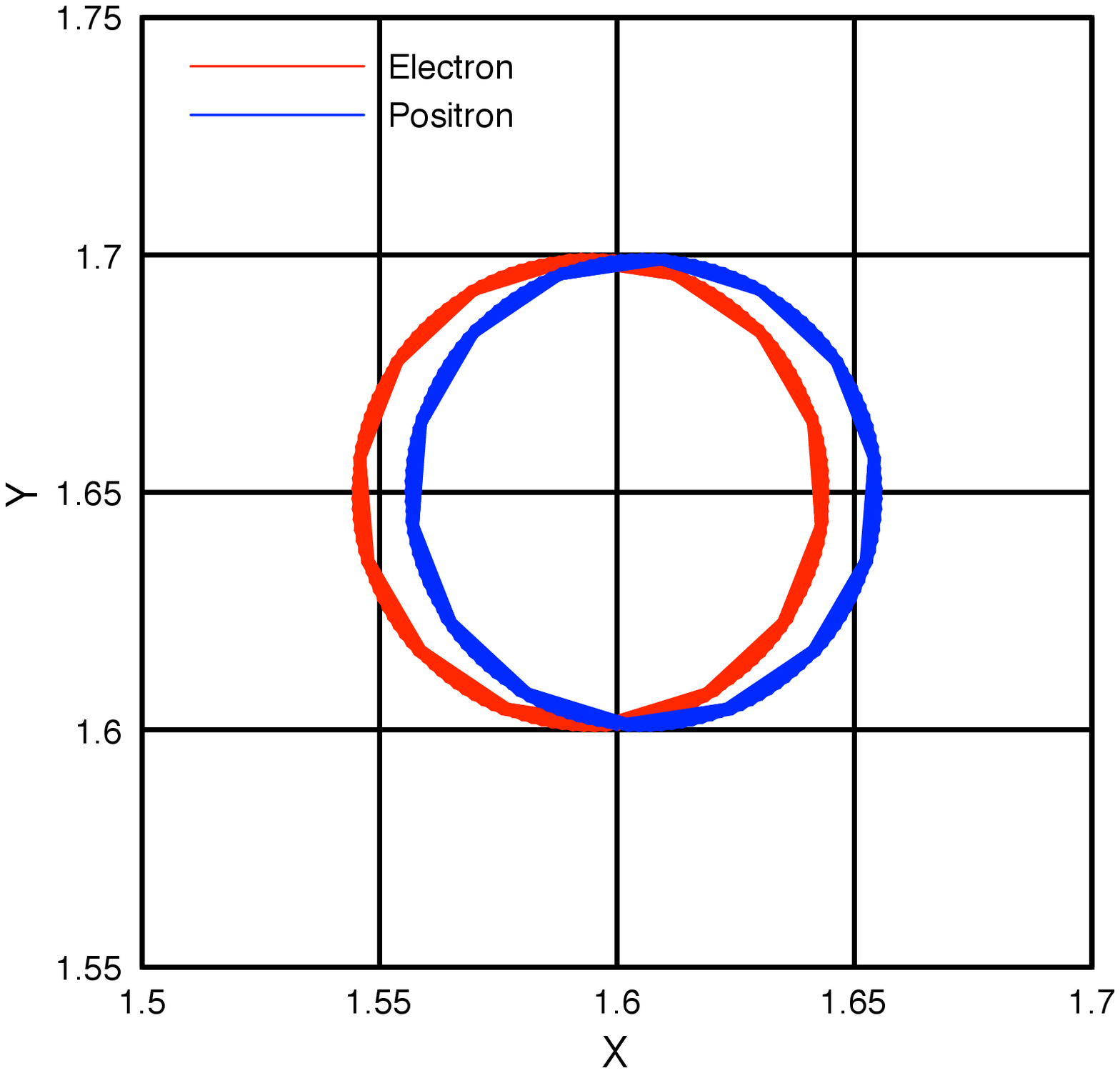}
\caption{Test results with increased particle charge. Left panel:
  particle energy and longitudinal electric field energy for the
  Langmuir oscillation. Right panel: electron and positron trajectories
in the magnetic field, black lines show the grid (cell boundaries).}
\end{figure}

Our comparison with the field interpolation
scheme presented in Eq.(\ref{eq:wrong1}) does not aim to criticize
\cite{buneman2,barth}, moreover that Eq.(\ref{eq:wrong1}) in TRISTAN is
used in a combination with low order $l=1$ form-factor for the charge
density, so that reducing this order following Eq.(\ref{alphax}). would be
impractical and still would not allow us to reach the energy
conservation. However, with the higher order form-factor, as we see,
the development of the interpolation scheme should go along the
direction different from that presented in Eq.(\ref{eq:wrong1}). 
\section{Conclusion}
The pulsed electric field in a focus of a high-field laser may
cause a charge separation even in the target of a
solid-state density. To simulate the 
laser-plasma interaction in that strong fields, the 
ChCPIC scheme could be the best choice (see \cite{esirkepov}). However,
these schemes are believed to be too noisy and the energy non-conservation, as we see,
can potentially be a source for such noise. 

We show that the alternating-order form-factor is to
mitigate this inherent flaw.

I am grateful to Prof. T. Zh. Esirkepov, 
Prof. K. Powell and Dr. N. Naumova for discussions.
\section*{Appendix. Full algorithm to advance the PIC solution.}
Assume the magnetic 
field and the particle momenta to be known at $t=n-1/2$, as well as the electric 
field and particle positions at 
$t=n$. Advance all the fields and particle through a time step. First,
advance the 
magnetic 
field through a half time step:
$$
B^{(x)\ n}_{i+1/2,j,k}=B^{(x)\ n-1/2}_{i+1/2,j,k}-
$$
\begin{equation}\label{updateB}
-\frac{c_y}{2}(E^{(z)\ n}_{i+1/2,j+1/2,k}-E^{(z)\ n}_{i+1/2,j-1/2,k})+
\frac{c_z}{2}(E^{(y)\ n}_{i+1/2,j,k+1/2}-E^{(y)\ n}_{i+1/2,j,k-1/2}),...
\end{equation}
Alternatively, the vector potential may be updated through a half time
step.
 
Then the particle motion is updated. For each particle,  
first, the fields at the particle position 
should be interpolated using Eq.(\ref{alphax}) and either Eq.(\ref{betax}) 
with the updated magnetic field
or Eq.(\ref{viaA}) with the update vector potential. To do
this, the alternating-order form-factor should be calculated for the
particle position, ${\bf x}_p^{n}$. Then, the particle 
momentum is advanced through a half time step, accounting for the 
effect from the electric field only:
$$
\tilde{\bf w}^{n}_p={\bf w}^{n-1/2}_p+{\bf e}({\bf x}^{n}_p),\qquad 
{\bf e}({\bf x}^{n}_p)=\frac{q_p\Delta t}{2m_p}{\bf E}({\bf x}^{n}_p),
$$
Then, the contribution from the magnetic force is added:
$$
{\bf w}^{n}_p=\tilde{\bf w}^{n}_p+
[\tilde{\bf w}^{n}_p\times{\bf b}({\bf x}^{n}_p)],\qquad 
{\bf b}({\bf x}^{n}_p)=\frac{q_p\Delta t}{2m_p\sqrt{(\tilde{\bf w}^{n}_p)^2+c^2}}
{\bf B}({\bf x}^{n}_p).
$$
The momentum is advanced through the full time step then:
$$
{\bf w}^{n+1/2}_p=\tilde{\bf w}^{n}_p+{\bf e}({\bf x}^{n}_p)+
\frac{2[{\bf w}^{n}_p\times{\bf b}({\bf x}^{n}_p)]}
{1+({\bf  b}^2({\bf x}^{n}_p))},
$$
the correction in the last term is chosen in such a manner that the magnetic 
field does not affect the particle energy, so that
\begin{equation}\label{p2}
({\bf w}^{n+1/2}_p-{\bf e}({\bf x}^{n}_p))^2=({\bf w}^{n-1/2}_p+{\bf e}({\bf x}^{n}_p))^2.
\end{equation}

Finally, the particle current should be calculated. First, save the form-factor at the cell centers,
$\Delta F_{i+1/2}(x_p^n)$, $\Delta F_{j+1/2}(y_p^n)$, $\Delta
F_{k+1/2}(z_p^n)$. Then calculate
the particle velocity and new particle position: 
\begin{equation}\label{coordscheme}
\frac{{\bf u}^{n+1/2}_p}c=\frac{{\bf w}^{n+1/2}_p}{\sqrt{\left({\bf w}^{n+1/2}_p\right)^2+c^2}},\qquad{\bf x}^{n+1}_p={\bf x}^{n}_p+
\frac{{\bf u}^{n+1/2}_p}c\cdot{\bf diag}(c_x,c_y,c_z)
\end{equation}
Find the new form-factors, $\Delta F_{i+1/2}(x_p^{n+1})$, 
$\Delta F_{j+1/2}(y_p^{n+1})$, $\Delta F_{k+1/2}(z_p^{n+1})$,
then calculate $\Delta F^\pm_{i+1/2}$, 
$\Delta F^\pm_{j+1/2}$, $\Delta F^\pm_{k+1/2}$ and, on calculating the
partial sums of
$\Delta F^-_{i+1/2}$, 
$\Delta F^-_{j+1/2}$, $\Delta F^-_{k+1/2}$, find the currents, $\xi$, using 
Eq.(\ref{current}). 
End update for this particle and proceed to the next one.

The magnetic field then should be advanced through another half time
in the way as presented in Eq.(\ref{updateB}). 
Finally the electric field should be updated, with the electric current density, which
sums up the contributions 
as in Eq.(\ref{current}) from all particles:
$$
{E}^{(x)\ n+1}_{i,j+1/2,k+1/2}=E^{(x)\ n}_{i,j+1/2,k+1/2}-4\pi
\Delta t 
J^{(x)\ n+1/2}_{i,j+1/2,k+1/2} +
$$
\begin{equation}\label{escheme}
+
c_y(B^{(z)\ n+1/2}_{i,j+1,k+1/2}-B^{(z)\ n+1/2}_{i,j,k+1/2})-
c_z(B^{(y)\ n+1/2}_{i,j+1/2,k+1}-B^{(y)\ n+1/2}_{i,j+1/2,k}),...
\end{equation}
End update for this time step and proceed to the next one.
 
\end{document}